\documentclass[12pt,a4paper]{article}
\pdfoutput=1
\usepackage{amsmath,amssymb}
\usepackage[pdftex]{graphicx}
\usepackage[a4paper, left=2cm, right=2cm, top=2cm, bottom=2cm]{geometry}
\newcommand{\pdiff}[3][]{\dfrac{\partial^{#1} #2}{\partial {#3}^{#1}}}

\title{Viscous flows in corner regions: Singularities and hidden eigensolutions}

\author{James E. Sprittles and Yulii D. Shikhmurzaev}

\begin{document}
\maketitle

\begin{abstract}
Numerical issues arising in computations of viscous flows in corners formed by a liquid-fluid free
surface and a solid boundary are considered. It is shown that on the solid a Dirichlet boundary
condition, which removes multivaluedness of velocity in the `moving contact-line problem' and gives
rise to a logarithmic singularity of pressure, requires a certain modification of the standard
finite-element method. This modification appears to be insufficient above a certain critical value
of the corner angle where the numerical solution becomes mesh-dependent. As shown, this is due to
an eigensolution, which exists for all angles and becomes dominant for the supercritical ones. A
method of incorporating the eigensolution into the numerical method is described that makes
numerical results mesh-independent again. Some implications of the unavoidable finiteness of the
mesh size in practical applications of the finite-element method in the context of the present
problem are discussed.

\end{abstract}

\section{Introduction}

The ability of a numerical scheme to accurately approximate a physical problem in a domain
containing corners is critical for the description of a number of phenomena, ranging from
electromagnetic wave propagation in a waveguide \cite{juntunen00} to die-swell effects in polymer
extrusion \cite{georgiou90}. Often, an analysis of such problems reveals singular behaviour of
variables as the corner is approached, which requires special numerical treatment. Such problems
are well known and have been thoroughly investigated in the setting of fracture mechanics, where
one considers the propagation of a crack into a material \cite{duflot06}, and have also been
studied in some fluid dynamics problems \cite{wilson06}.

Our interest here is the viscous flow in a corner formed between a liquid-fluid free surface and a
solid boundary. Although a free surface is generally bent, to leading order as the corner is
approached, it is often possible for the purpose of a local analysis to consider the flow domain as
having a wedge shape. This is the case, for example, in dynamic wetting flows \cite{shik07}, where
the free surface and the solid boundary form what is referred to as the `contact angle' and the
liquid-fluid-solid `contact line' moves with respect to the solid surface. This differs from the
situation considered in some previous investigations on flow in a corner where the contact line is
stationary with respect to the solid and motion is generated by disturbances in the far field
\cite{georgiou89,georgiou90}.

It is well known that the classical fluid-mechanical approach when applied to dynamic wetting
problems fails to provide an adequate description of the flow \cite{huh71}.  The conventional
remedy to the problem is to relax the no-slip boundary condition on the solid surface and allow for
`slip' between the liquid and solid. A number of different forms for this slip behaviour have been
examined in the literature (for a recent review see Ch.~3 of \cite{shik07}). Broadly, these split
into conditions which (i) relate tangential stress to the slip velocity, such as the Navier
condition \cite{navier23}, or (ii) explicitly prescribe the velocity along the solid surface.

In this paper, we consider numerical problems arising in the second case where the velocity along
the solid surface is \emph{a priori} prescribed in a form which ensures that a solution exists, and
that in the far field the usual no-slip condition is restored. This approach is appealing to some
users due to its mathematical simplicity, and our goal here is to show numerical pitfalls one comes
across in its numerical implementation and give a method of overcoming them which provides a
framework for modelling this class of problems. A number of functions prescribing the fluid
velocity on the solid surface have been proposed in the literature
\cite{dussan76,zhou90,somalinga00} and here we consider just one of these which captures all the
main features of the problem.

\section{Problem formulation}\label{pf}

The problem is most easily formulated in a polar coordinate system $(r,\theta)$ in a frame moving
with the contact line (now referred to in our two-dimensional domain as the corner point). The
wedge is formed by a solid surface at $\theta=0$ which moves at speed $U$ parallel to itself, a
flat free surface at $\theta=\alpha$ and a `far field' boundary which is placed at an arc of a
sufficiently large radius $r=R$.

The liquid is Newtonian and incompressible, with density $\rho$ and viscosity $\mu$. Near the
corner the flow is characterized by a small length scale so that the Reynolds number $\hbox{\it
Re\/}$ based on this scale is small. Then as $\hbox{\it Re\/}\to0$, to leading order in $\hbox{\it
Re\/}$  we have the Stokes flow\footnote{The analysis remains valid for the full Navier-Stokes
problem since it considers the $r\to0$ limit, and here we consider the Stokes equations in the
whole region as a convenient way to illustrate the idea.}. The non-dimensional Stokes equations for
the bulk pressure $p$ and the radial and azimuthal components of velocity $(u,v)$ take the form:
\begin{equation}
\label{contin} \frac{1}{r}\frac{\partial(ru)}{\partial r}
+\frac{1}{r}\frac{\partial v}{\partial\theta}=0,
\qquad\qquad(0<r<R,\ 0<\theta<\alpha),
\end{equation}

\begin{equation}
\label{motion_prim}
\frac{\partial p}{\partial r}=\Delta u
-\frac{u}{r^2}-\frac{2}{r^2}\frac{\partial v}{\partial\theta},
\qquad \frac{1}{r}\frac{\partial p}{\partial\theta}=\Delta v
-\frac{v}{r^2}+\frac{2}{r^2}\frac{\partial u}{\partial\theta},
\end{equation}
where
$$
\Delta=\frac{\partial^2}{\partial r^2}+\frac{1}{r}\frac{\partial
}{\partial r} +\frac{1}{r^2}\frac{\partial^2}{\partial\theta^2}.
$$

On the solid surface, for a solution not to have multivalued velocity at the corner point
\cite{dussan74,shik07}, we replace the no-slip condition ($u=1$) with a prescribed velocity that
has free-slip at the corner point and attains no-slip in the far field, that is:
\begin{equation}
\label{form} u = 0,\quad\hbox{at}\quad r=0\qquad\hbox{and}\qquad u\rightarrow1,\quad \hbox{as}\quad
r\rightarrow \infty.
\end{equation}
Following \cite{somalinga00}, we use an exponential form for this function and, to complete the
boundary conditions on the solid surface, it is combined with the usual impermeability condition
for the component of velocity normal to the surface:
\begin{equation}
\label{vect_ss} u = 1-\exp(-r/s),\quad v=0, \qquad\qquad(0<r<R,\
\theta=0).
\end{equation}
The region in which the velocity deviates from no-slip is characterized by the value of $s$, which
is a (non-dimensional) `slip length'.

On the free surface, we have the standard boundary conditions of zero tangential stress and
impermeability:
\begin{equation}
\label{vect_fs} \frac{\partial u}{\partial\theta}=0,\quad
v=0,\qquad\qquad(0<r<R,\ \theta=\alpha).
\end{equation}

In the far field, we assume that the flow is fully developed and apply `soft' conditions:
\begin{equation}
\label{vect_far_field} \frac{\partial u}{\partial r}=\frac{\partial
v}{\partial r}=0, \qquad\qquad(r=R,\ 0<\theta<\alpha),
\end{equation}
which imply that the influence of slip has attenuated; these conditions are satisfied by the
(multivalued at the corner point) solution obtained using the no-slip condition all along the solid
surface \cite{moffatt64}.

Equations (\ref{contin})--(\ref{vect_far_field}) fully specify the
problem of interest.

\section{Local asymptotics}\label{ana}

Consider the leading-order asymptotics for the solution of (\ref{contin})--(\ref{vect_far_field})
as $r\to0$ \cite{shik07} that we will later need to use in the numerical code and to provide a test
of accuracy of the numerical results presented in the next section. After introducing the stream
function $\psi$ by
\begin{equation}
\label{streamfunction} u = \frac{1}{r}\frac{\partial \psi}{\partial \theta}, \quad v = -
\frac{\partial \psi}{\partial r},
\end{equation}
equations (\ref{contin})--(\ref{motion_prim}) are reduced to a biharmonic equation $\Delta^2\psi=0$
with boundary conditions (\ref{vect_ss})--(\ref{vect_fs}) taking the form
\begin{equation}
\label{polar_slip} \frac{\partial\psi}{\partial\theta} =
r\left(1-\exp\left(-r/s\right)\right), \quad\psi=0,
\qquad\qquad(\theta=0,\ 0<r<R),
\end{equation}
\begin{equation}
\label{polar_free_slip}
\frac{\partial^{2}\psi}{\partial\theta^{2}}=0, \quad \psi=0,
\qquad\qquad(\theta=\alpha,\ 0<r<R),
\end{equation}
where, for definiteness, we assign the value zero to the streamline coinciding with the wedge's
boundary.

Condition (\ref{polar_slip}) is the only inhomogeneous boundary condition in the problem, i.e.\ the
condition that drives the flow. To leading order as $r\to0$, it has the form
\begin{equation}\label{u_form}
\left.\pdiff{\psi}{\theta}\right|_{\theta=0}=ar^{2}+O\left(r^{3}\right),
\end{equation}
where $a=1/s$. An alternative prescribed velocity that satisfies (\ref{form}) and (\ref{u_form})
known in the literature \cite{dussan76} is given by $u=(r/s)/(1+r/s)$ and the asymptotic analyses
throughout this paper are equally valid for this function as well. The form (\ref{u_form}) suggests
looking for the leading-order term of the local asymptotics in the form $\psi=r^2F(\theta)$, which
is a particular case from a known family of separable solutions of the biharmonic equation of the
form $\psi=r^\lambda F(\theta)$. After substituting $\psi=r^2F(\theta)$ into $\Delta^2\psi=0$, one
arrives at
\begin{equation}
\label{stream_2} \psi = r^{2}
\left(B_{1}+B_{2}\theta+B_{3}\sin2\theta+B_{4}\cos2\theta\right),
\end{equation}
where the constants of integration $B_i~(i=1,\dots,4)$, found from
(\ref{polar_slip})--(\ref{polar_free_slip}), are given by\footnote{Here, we correct a typographical
error in $B_1$ on p.~126 of \cite{shik06} and on p.~153 of \cite{shik07}.}:
\begin{equation}\label{constants}
B_1=-B_4=\frac{a\alpha\sin2\alpha}{2\alpha\cos2\alpha-\sin2\alpha},\qquad B_2=-B_1/\alpha,\qquad
B_3=B_1\cot2\alpha.
\end{equation}

The pressure field obtained from (\ref{motion_prim}) using (\ref{streamfunction}) and
(\ref{stream_2}) has the form
\begin{equation}\label{pressure}
p=4B_{2}\ln r + p_{0}.
\end{equation}
where $p_{0}$ is a constant which sets the pressure level.

It is immediately obvious from (\ref{constants}) that the coefficients are singular when
$2\alpha\cos2\alpha-\sin2\alpha=0$, which occurs at a critical value $\alpha_{c}$ determined by
$\tan(2\alpha_{c})=2\alpha_{c}$. In the range of interest, i.e.\ for $0<\alpha_{c}<180^\circ$, we
have $\alpha_{c}\approx128.7^\circ$.

It is noteworthy, that in the limit $r\to0$, the velocity scales {\it linearly\/} with $r$ whilst
the pressure is {\it logarithmically\/} singular at the corner and is independent of the angular
coordinate $\theta$.

\section{Numerical results}\label{num}

From a numerical viewpoint, the steady fixed-boundary problem considered in this paper is
complicated only by the presence of a singularity in the pressure which, according to
(\ref{pressure}), is logarithmic as $r\rightarrow 0$. The simplicity of the rest of the problem and
the availability of asymptotic results, which not only give the behaviour of the velocity and
pressure near the corner, but also provide the coefficients, make this a perfect testing ground for
a numerical method's ability to approximate flows in corner regions formed by boundaries on which
different types of boundary conditions are applied.

In the standard implementation of the finite-element, as well as finite-difference, algorithm, one
assigns an {\it a priori\/} unknown finite value to the pressure at the corner point. If such a
code attempts to approximate a solution where the pressure at the corner point is singular, like
the one whose local asymptotics we considered earlier, the nodal value of the pressure, as well as
the pressure at the neighbouring nodes, will vary as one refines the mesh. In other words, an
attempt to approximate a singular analytic solution using regular numerical representations of the
unknown function on each element will lead to a numerical `solution' that is mesh-dependent and
hence, strictly speaking, it is not a solution to the original problem formulated in terms of PDEs
that `do not know' about any mesh.

In order to achieve a uniformly valid solution in the framework of a finite-element method, one
approach is to use singular elements, i.e.\ to redefine the pressure interpolation in the elements
that contain the corner point node in such a way that the pressure is allowed to be infinite at the
corner point and behave as described by the local asymptotics. A simple implementation of this idea
is described in \cite{wilson06}.

In the present work, the problem formulated in Section~\ref{pf} has been considered using part of a
finite-element-based numerical platform which has been developed to simulate a range of
microfluidic capillary flows and has already been used to obtain new results for the flow of
liquids over surfaces of varying wettability \cite{sprittles07,sprittles09}. The idea here, is to
modify the finite-element's basis function $\Phi_p$ associated with the pressure at the corner
point. In the standard triangular Taylor-Hood element with six velocity nodes and three pressure
nodes, $\Phi_p$ is linear and takes the value 1 at the corner point and 0 at all other pressure
nodes. Now, instead of using $\Phi_p$ and determining the coefficient in front of it, we will be
using and determining the coefficient in front of a singular basis function $\Phi_s=\Phi_p \ln r$.
We denote the computed coefficient of $\Phi_s$ as $B_p$. Instead of $\Phi_p$ one could use other
functions to pre-multiply the logarithm, for example $\sin(\pi\Phi_p/2)$; such functions have been
tried and it was found that they do not offer noticeable advantages over $\Phi_p$.

It is pointed out in \cite{wilson06} that the usual Gaussian quadrature is not well suited to the
integration of a singular function, such as $\ln r$, and a special Simpson quadrature routine has
been suggested to provide a very accurate approximation of the integrals. This routine has also
been incorporated into our numerical platform; however, we found that using Gaussian quadrature
with enough integration points provided an accurate enough estimation of the integral and was
significantly quicker and easier to implement.  Sixteen Gauss integration points were found to be
more than sufficient.

At the critical angle $\alpha_{c}\approx127.8^\circ$, more complex asymptotic analysis should be
considered to resolve the singular behaviour and the results should be incorporated into the code
in a way similar to what is being described. However, since this paper is concerned with the
general numerical approximation of corner flows which contain singularities and their numerical
treatment, we shall consider angles away from this critical value, i.e.\ the range where, as one
might expect at this stage, our asymptotics of Section~\ref{ana} can be used.  First, we shall
consider subcritical angles $\alpha<\alpha_{c}$, in particular $\alpha=75^\circ$ as a
representative case.  We take $s=0.1$ and $R=10$ in all the simulations that we present as an
investigation into the variation of these parameters would be about the physical problem rather
than its numerics and would detract from the main emphasis of this paper.

\subsection{Numerical approximation at subcritical corner angles}\label{75}

\begin{figure}
\centering
\includegraphics*[scale=0.6]{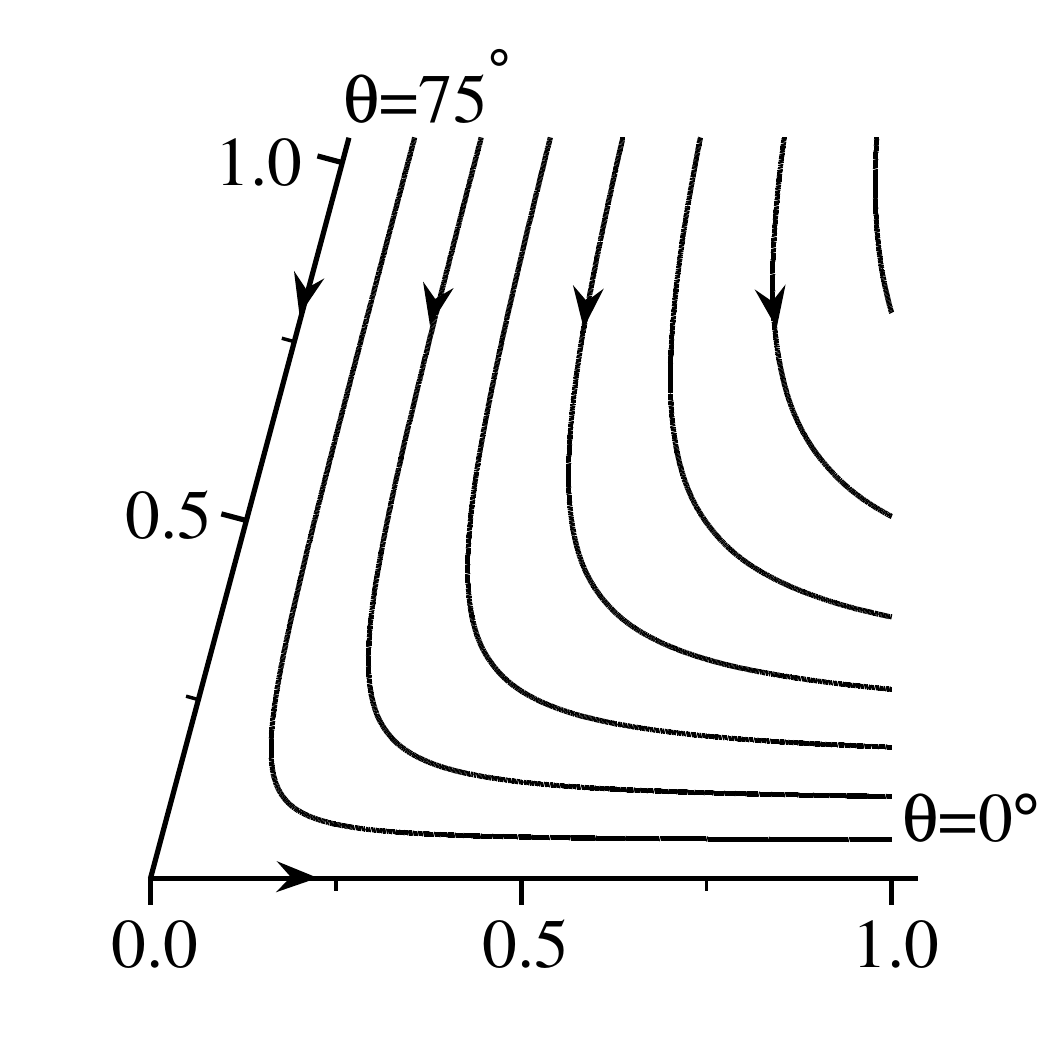}
\includegraphics*[scale=0.6]{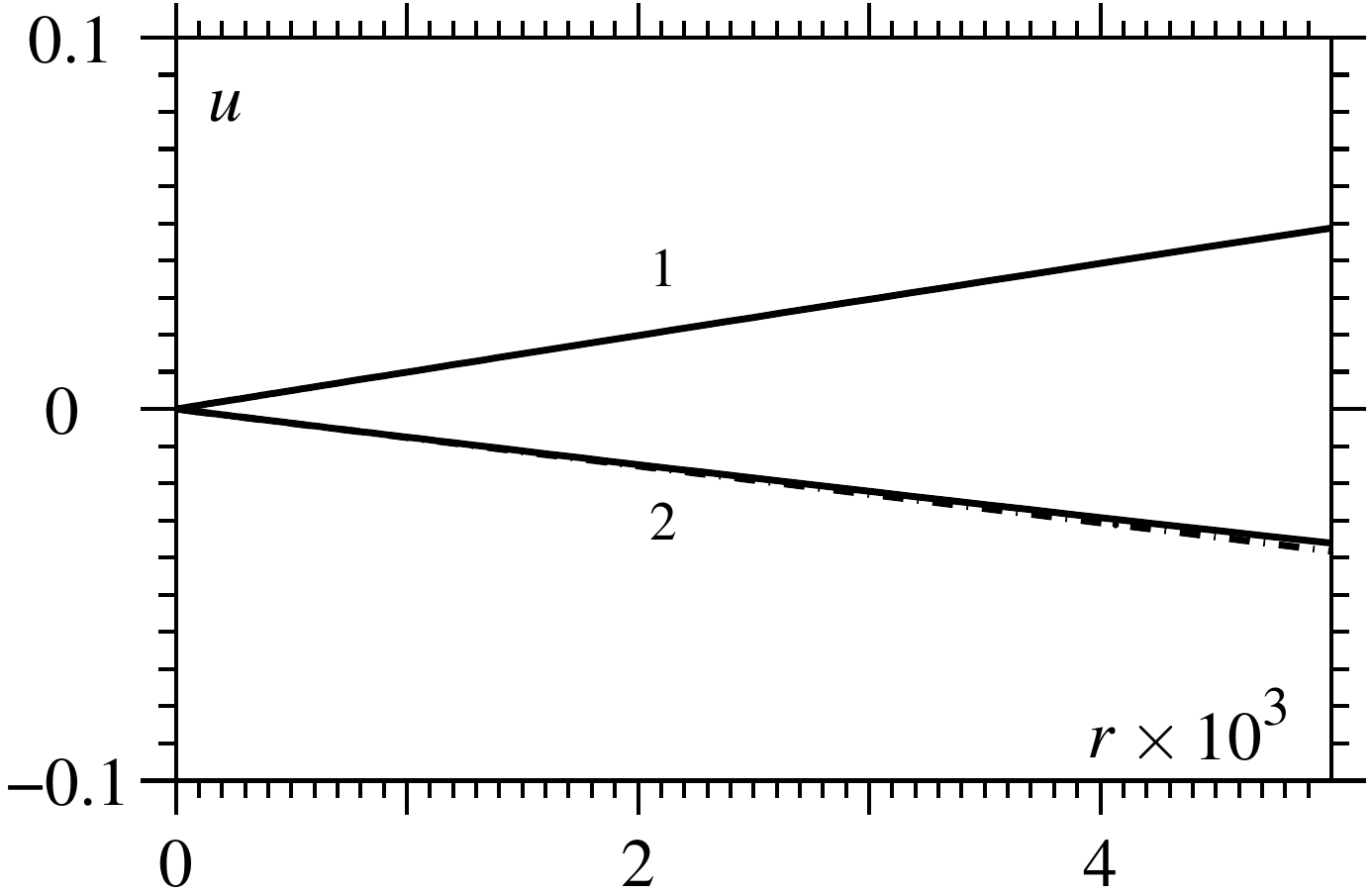}
\caption{Left: streamlines in increments of $\psi=0.05$, with the $\theta=0,\alpha$ ~interfaces
corresponding to $\psi=0$. Right: comparison of the computed velocity with the analytic prediction
(dashed line) along the liquid-solid (1) and liquid-fluid (2) interfaces.}\label{F:75velocity}
\end{figure}

In Fig.~\ref{F:75velocity}, we show the streamlines generated by the exponential slip model.  As
expected, the prescribed velocity on the solid draws fluid near the solid out of the corner, thus
reducing the pressure there which then sucks in fluid from the far field. In the same figure, the
components of the radial velocity along the interfaces are compared to the asymptotic predictions;
the agreement between the calculated velocity along the liquid-fluid interface and the asymptotic
prediction is visibly excellent (agreement along the liquid-solid interface near the corner is
guaranteed as the velocity is prescribed, so we give it here just to show the range in which the
velocity varies).

\begin{figure}
\centering
\includegraphics*[scale=0.55]{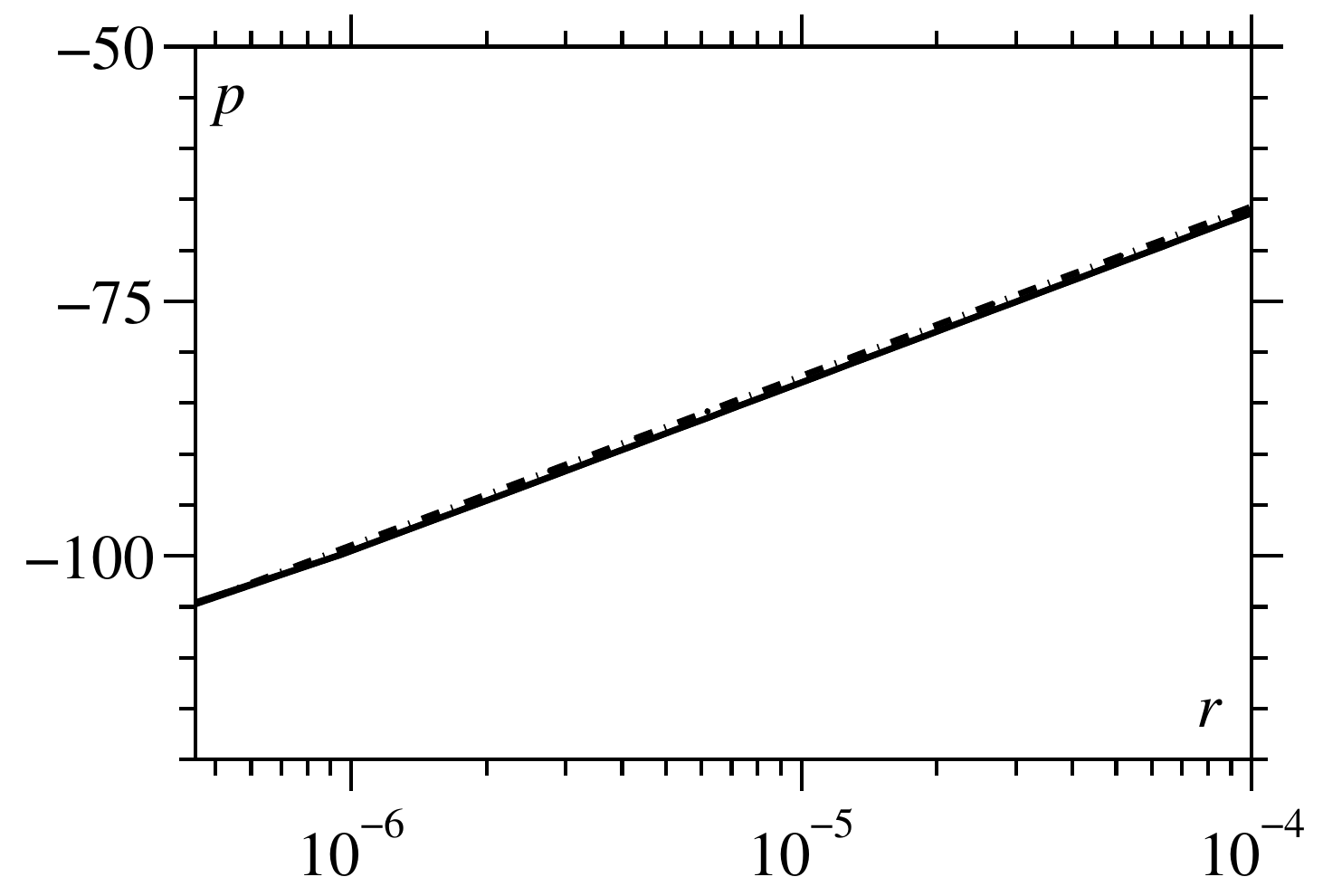}
\includegraphics*[scale=0.62]{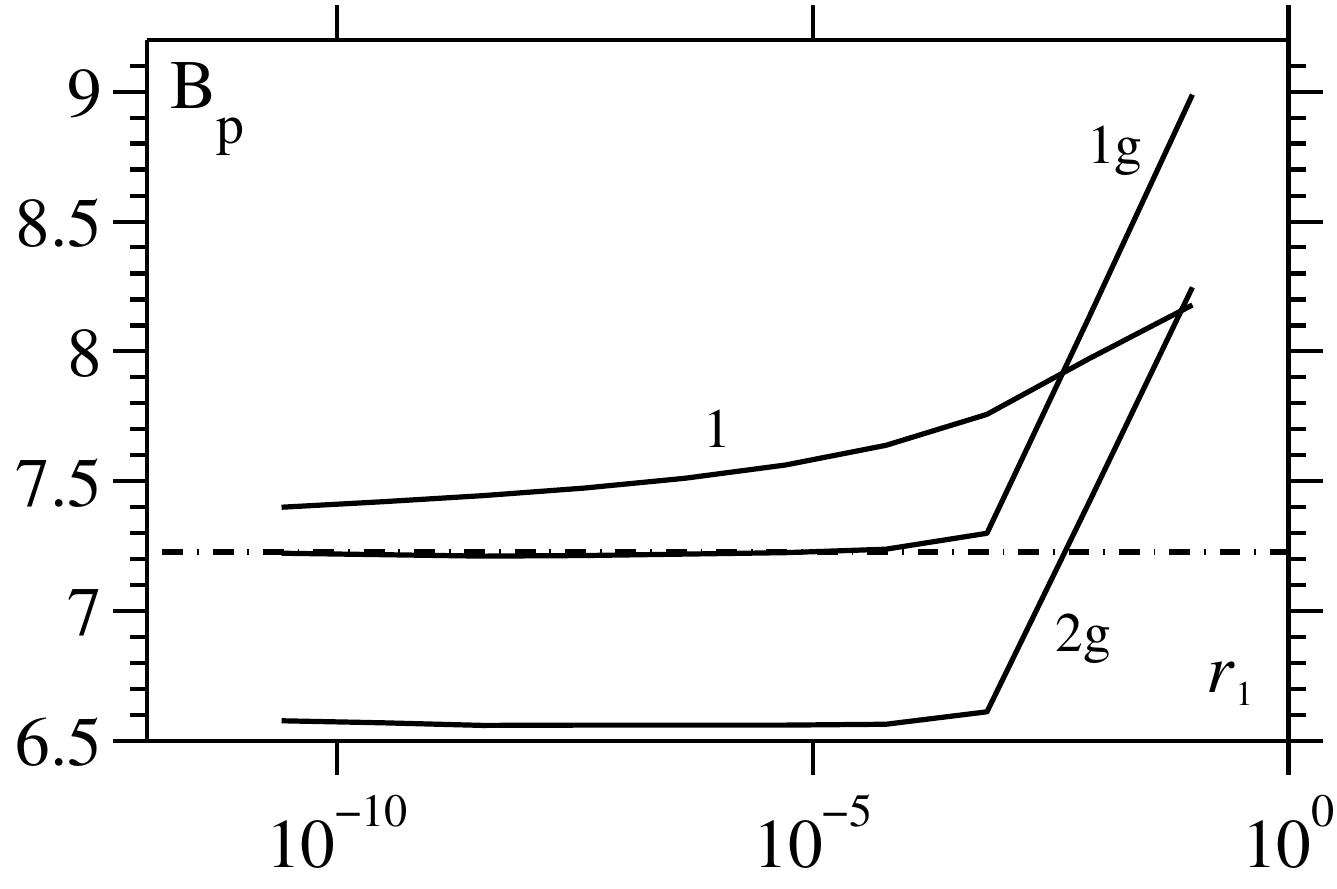}
\caption{Left: comparison of computed pressure along the $\theta=0,\alpha$ ~interfaces, which are
graphically indistinguishable, as the corner is approached with the analytic pressure (dashed
line). Right: convergence of the pressure coefficient of $\ln r$ with (1) and without (2) singular
elements as the mesh is resolved. $B_{p}$ is calculated from the singular element's coefficient (1)
and from the local gradient (1g) and (2g). The analytic value is 7.23 and $r_{1}$ is the size of
the smallest element. }\label{F:75pressure}
\end{figure}

The plot of pressure along the two interfaces in Fig.~\ref{F:75pressure} shows that the pressure is
indeed $\theta$-independent, and it is almost graphically indistinguishable from the asymptotic
result. To confirm the mesh-independence of the result, we also consider how well the singular
behaviour of pressure is captured as the mesh is resolved over ten orders of magnitude using ten
different meshes, characterized by the width of the smallest element $r_1$. To do so, we show both
the coefficient $B_p$ in front of the basis function $\Phi_s$ (curve 1 in Fig.~\ref{F:75pressure})
and the appropriate gradient determined from the pressures $p_1,\ p_2$ at the two pressure nodes on
the solid surface, for simplicity, closest to the corner point at radial distances $r_1,\ r_2$,
which is given by $(p_{2}-p_{1})/(\ln r_{2} - \ln r_{1})$ (curve~1g). This second method allows us
to compare the code with the singular elements to one without (curve~2g in the plot) where $B_p$ is
not explicitly calculated.

The clearest conclusions are drawn from the results of the second method (curves 1g and 2g). Here,
the plot shows that, by using the singular element, the local gradient converges to the asymptotic
value of $B_{p}=7.23$ (the dashed line in the figure): without these singular elements the code
converges to the incorrect value as the mesh is resolved. This is strong support for the inclusion
of singular elements in dynamic wetting codes. Without them, the behaviour of pressure is wrong not
only in the element adjacent to the corner point, but, by continuity, also in a neighbourhood of
this element.

For $\alpha<\alpha_{c}$ the asymptotics and numerics are in excellent agreement,  and the special
treatment of the corner singularity was a success.  Now  we consider a supercritical angle
$\alpha=175^\circ>\alpha_{c}$.

\subsection{Numerical approximation at supercritical corner angles}

\begin{figure}
\includegraphics*[angle=0,scale=0.45]{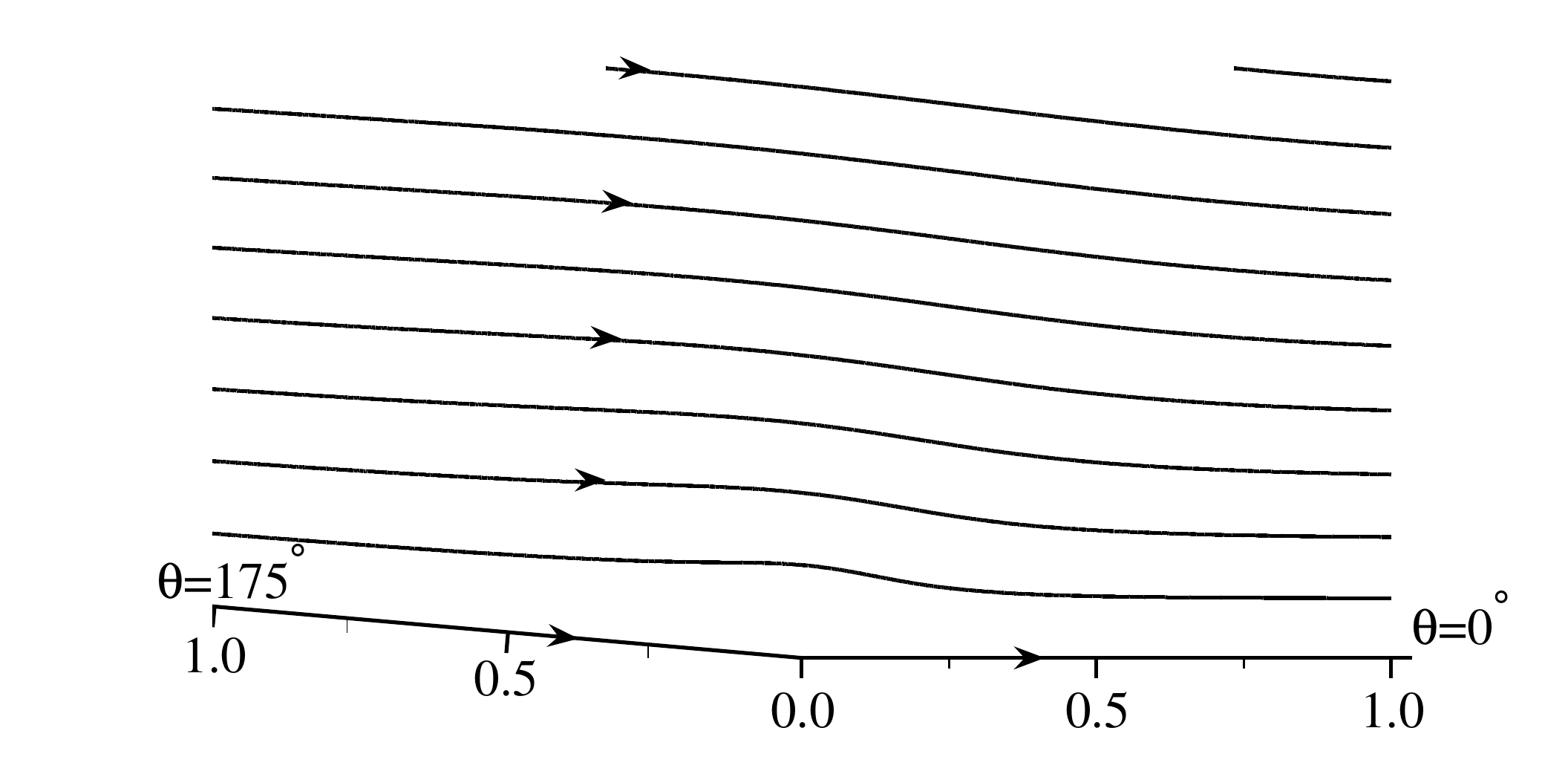}
\includegraphics*[scale=0.55]{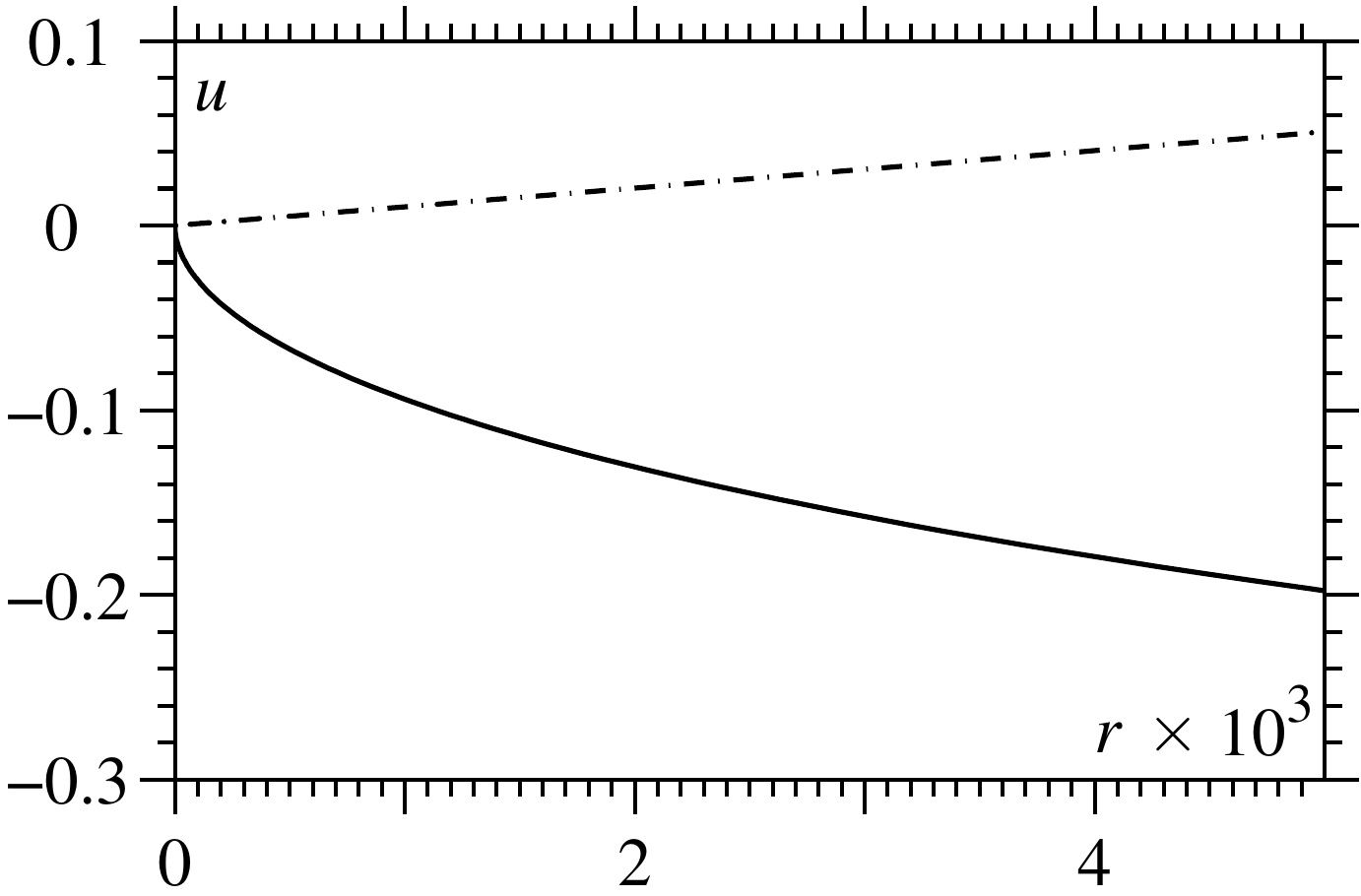}
\caption{Left: streamlines in increments of $\psi=0.1$ with the $\theta=0,\alpha$ ~interfaces
corresponding to $\psi=0$. Right: comparison of the velocity along the liquid-fluid interface with
the analytic prediction (dashed line).}\label{F:175velocity}
\end{figure}

The streamlines in Fig.~\ref{F:175velocity} are as one may intuitively expect, but when we compare
the numerical and asymptotic results for the velocity along the liquid-fluid interface there is no
agreement. In fact, the asymptotic result predicts that the flow should be {\it up\/} the
liquid-fluid interface, which is clearly not the case in the computed solution.

\begin{figure}
\centering
\includegraphics*[scale=0.55]{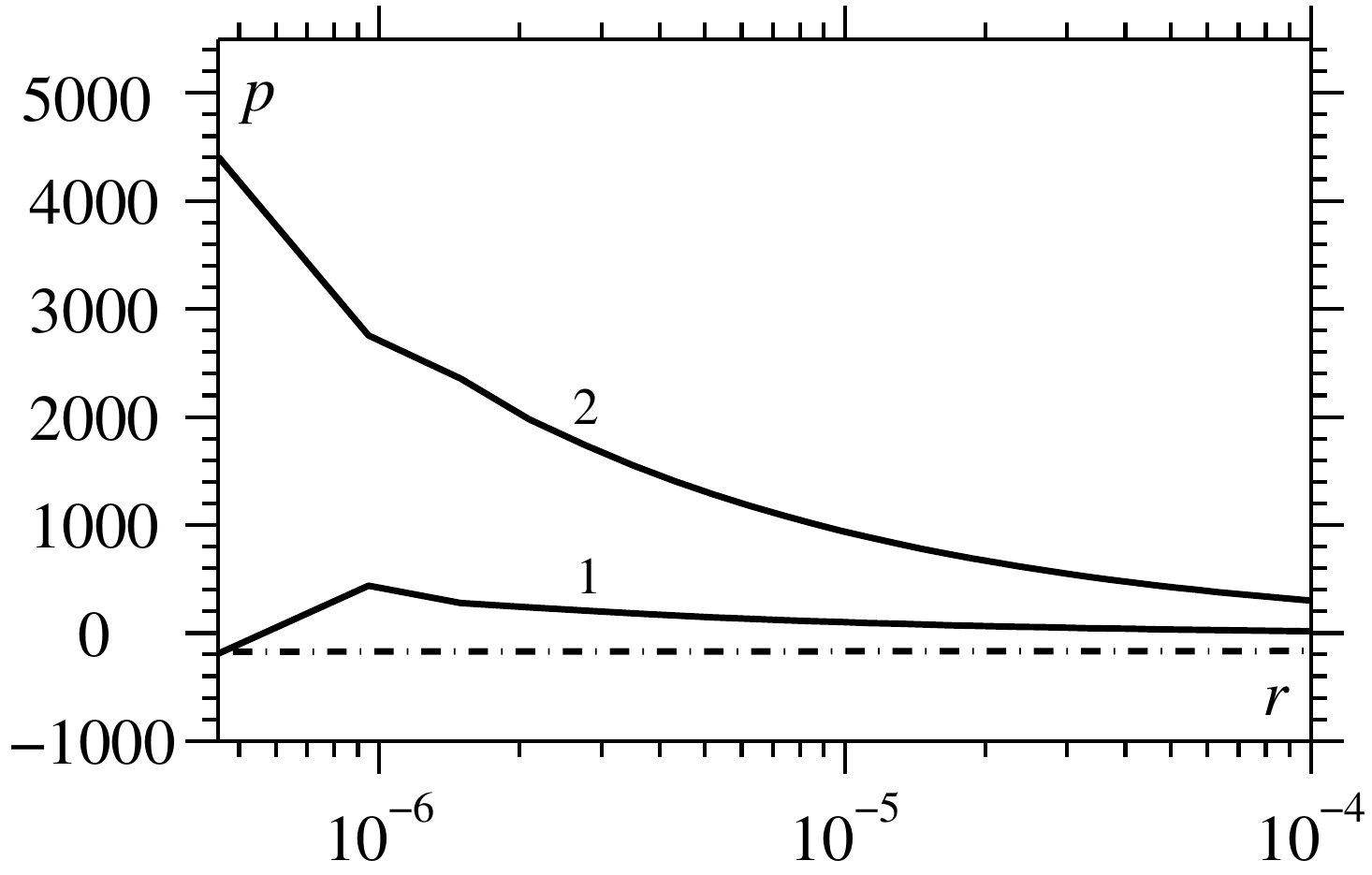}
\includegraphics*[scale=0.6]{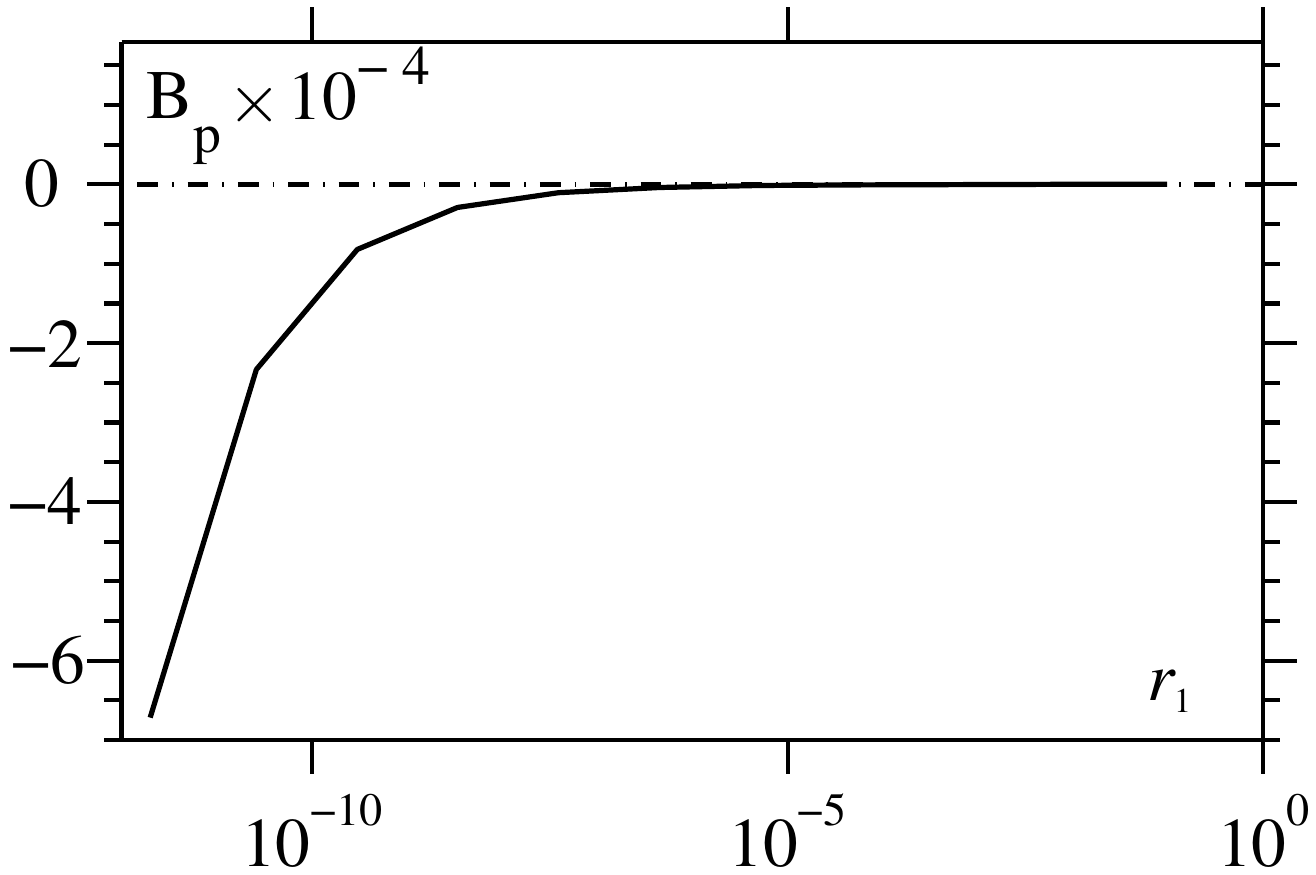}
\caption{Left: computed pressure in the vicinity of the corner along the liquid-solid (1) and
liquid-fluid (2) interfaces compared to the analytic prediction (dashed line). Right: pressure
coefficient compared to the analytic value 1.122 (dashed line) plotted against $r_{1}$ the size of
the smallest element. }\label{F:175pressure}
\end{figure}

The computed pressure along the liquid-solid interface is given as curve $1$ in
Fig.~\ref{F:175pressure}. It is not only that the pressure strongly deviates from the analytic
result (dashed line) as the corner point is approached; one can see that there also appear huge
oscillations: the pressure decrease in the element adjacent to the special corner elements (the
line to the left of the point $10^{-6}$ in the plot) is followed by a steep increase in the element
comprising the corner point (not shown in the semilogarithmic plot). Such mesh-dependence of the
numerical result indicates that the obtained solution cannot be regarded as a valid approximation
of the solution to the original set of partial differential equations. This conclusion is
re-enforced when we study the value of $B_{p}$, the coefficient to the logarithm in the singular
element, as we refine the mesh: there is no convergence. A similar trend is observed if we study
$B_p$ using the local gradient method.  The same conclusions may be drawn for all supercritical
angles.

Thus, the standard FEM coupled with the local-asymptotics-based approximation of the pressure does
not allow one to obtain an acceptable numerical approximation for solutions of the Stokes equations
in a corner with a combination of Dirichlet and Neumann boundary conditions on the interfaces for
all angles greater than $127.8^\circ$.

The robustness of the obtained numerical `solution' suggests that there is a fundamental numerical
problem. It should be emphasized that we arrived at this difficulty just by varying the wedge angle
in a code that produces excellent results for smaller wedge angles, and then cannot provide any
mesh-independent solution after a critical angle. An immediate (tentative) explanation for this
situation is that, besides the analytical solution whose asymptotics has been considered in
Section~\ref{ana} and incorporated into the code, there exists a `local eigensolution', i.e.\ a
solution satisfying zero boundary conditions on the sides of the wedge, that becomes dominant for
the supercritical angles. We will now examine this conjecture.

\subsection{Asymptotics of an eigensolution}\label{eigen}

The near-field asymptotics of the eigensolution to the biharmonic equation satisfying conditions
\begin{equation}\label{eigen_bcs}
\psi = \pdiff{\psi}{\theta} = 0,\quad\hbox{on}\quad\theta=0, \qquad\hbox{and}\qquad \psi =
\pdiff[2]{\psi}{\theta} = 0,\quad\hbox{on}\quad\theta=\alpha,
\end{equation}
is given by
\begin{equation}
\psi_{e} = r^{\lambda}\left[A_{1}\sin\left(\lambda\theta\right) +
A_{2}\cos\left(\lambda\theta\right) + A_{3}\sin\left(\left(\lambda-2\right)\theta\right) +
A_{4}\cos\left(\left(\lambda-2\right)\theta\right)\right].
\end{equation}
Using (\ref{eigen_bcs}) and noting that $\lambda\neq 1,2$, we find that $\lambda$ is determined by
the equation:
\begin{equation}\label{lambda_eqn}
2\sin\left(\lambda\alpha\right)\cos\left(\left(\lambda-2\right)\alpha\right)=\lambda\sin\left(2\alpha\right).
\end{equation}
Defining the degree of freedom by $A\equiv A_{1}$, the boundary conditions (\ref{eigen_bcs}) give:
\begin{equation}\label{coeffs}
A_{2}=-A_{4} =
-A\frac{\lambda(\lambda-2)^{-1}\sin\left(\left(\lambda-2\right)\alpha\right)-\sin\left(\lambda\alpha\right)}
{\cos\left(\left(\lambda-2\right)\alpha\right)-\cos\left(\lambda\alpha\right)},\qquad
A_{3} = -A\frac{\lambda}{\lambda-2}.
\end{equation}
and the pressure has the form:
\begin{equation}\label{p_eigen}
p_{e} =
4A(\lambda-1)r^{\lambda-2}\left[a_{3}\cos((\lambda-2)\theta)-a_{4}\sin((\lambda-2)\theta)\right]+p_{1}
\end{equation}
where $a_i=A_i/A,~i=3,4$ and $p_{1}$ is a constant setting the pressure level. Promisingly,
equation (\ref{lambda_eqn}) has roots $\lambda\in(1,2)$ for $\alpha\in (\alpha_{c},180^\circ)$,
with $\lambda\to2$ as $\alpha\to\alpha_c$, $\lambda\to3/2$ as $\alpha\to180^\circ$ and $\lambda$ as
a function of $\alpha$ varying monotonically between these limiting values.

This eigensolution has been derived in a number of other works considering the flow in a corner
formed between an impermeable no-slip boundary and an impermeable, sometimes free, zero-tangential
stress boundary, e.g. in \cite{anderson93}; these are sometimes referred to as `stick-slip
phenomena' \cite{richardson70,georgiou89}. The difference between our problem and the
aforementioned flows with static corner points is that, unlike these situations where the flow is
driven by the far field, here the fluid motion is generated by the movement of the solid and
analytically this behaviour is captured in the asymptotics of Section~\ref{ana}. The eigensolution
comes on top of this solution and, in the near field, it `does not know' about the motion of the
solid, although, ultimately, it is the solid's motion that generates the flow in the far field that
gives rise to this solution. The eigensolution exists in the range of subcritical angles as well,
but there it is regular in all variables and therefore causes no problem for numerical
computations; it is only for $\alpha>\alpha_c$ that the eigensolution becomes both singular and
dominant.

For $\alpha<\alpha_c$ the pressure at the corner point can be referred to as single-valued: the
coefficient in front of the logarithm is independent of $\theta$. In contrast, the solution for
$\alpha>\alpha_c$ is manifestly multivalued as predicted by the eigensolution (\ref{p_eigen}) and
as seen numerically: if one takes a vicinity of the corner point, then, no matter how small this
vicinity is, there will be points which are equidistant from the corner point with an arbitrarily
large pressure difference. Here, being interested in the numerical side of the problem, we set
aside physical arguments that might arise in connection with the obtained solution.

The existence of an eigensolution and its dominance for $\alpha>\alpha_c$ suggests that our initial
attempt at computing the flow at large angles were flawed because, given the asymptotics of Section
~\ref{ana}, we assumed that the pressure scaled as $\ln r$ whereas in fact the most singular term
(i) has order $r^{-k}$ where $k=2-\lambda\in(0,1)$ and (ii) is dependent on $\theta$. This suggests
a generalisation of our approach: we need to incorporate the new singular behaviour into the
special elements adjacent to the corner point. Due to the presence of the eigensolution, we now
have an unknown constant $A$ in our asymptotics which will prevent us from comparing \emph{a
priori} determined analytic curves with our numerical results.  However, once $A$ is determined
numerically, we may use it to extrapolate the asymptotic behaviour outside the singular element in
which it is calculated, i.e. we may then compare, a now semi-analytic, asymptotic prediction to the
computed solution \emph{globally}.

An alternative method that we used to verify our singular element solution and do not describe in
detail here is to analytically remove the eigensolution, which is the cause of numerical
difficulties, prior to computation and then superimpose it back on after.  This approach that has
been shown to be essential for the simulation of flows using the Navier slip condition, a
Robin-type boundary condition, on the solid surface is described elsewhere \cite{sprittles09a}. In
more complex problems, where the corner is just one element, this method of removing the
eigensolution everywhere is overly complex and a local method should be used which removes the
eigensolution near to the corner; this method is also described in \cite{sprittles09a} using
examples of full scale dynamic wetting simulations.

\subsubsection{Modified singular elements}

In the limit as $r\to0$, the first two terms in the expansion of pressure are $p=
A_{p}r^{\lambda-2}g(\theta)+B_{p}\ln r$.  For $\alpha>\alpha_c$ both terms are singular. We will
begin by using only the leading order term in $r$ in our singular elements and will return to the
two-term approximation later. Taking the leading term, our new singular elements have a basis
function of the form:
\begin{equation}
\Phi_s = \left\{
  \begin{array}{ll}
    \Phi_p \ln r                  , & \hbox{$\alpha<\alpha_c$,} \\
    \Phi_p r^{\lambda-2} g(\theta), & \hbox{$\alpha>\alpha_c$,}
  \end{array}
\right.
\end{equation}
where the unknown coefficients are $B_{p}$ and $A_{p}$, respectively.

\begin{figure}
\centering
\includegraphics*[scale=0.55]{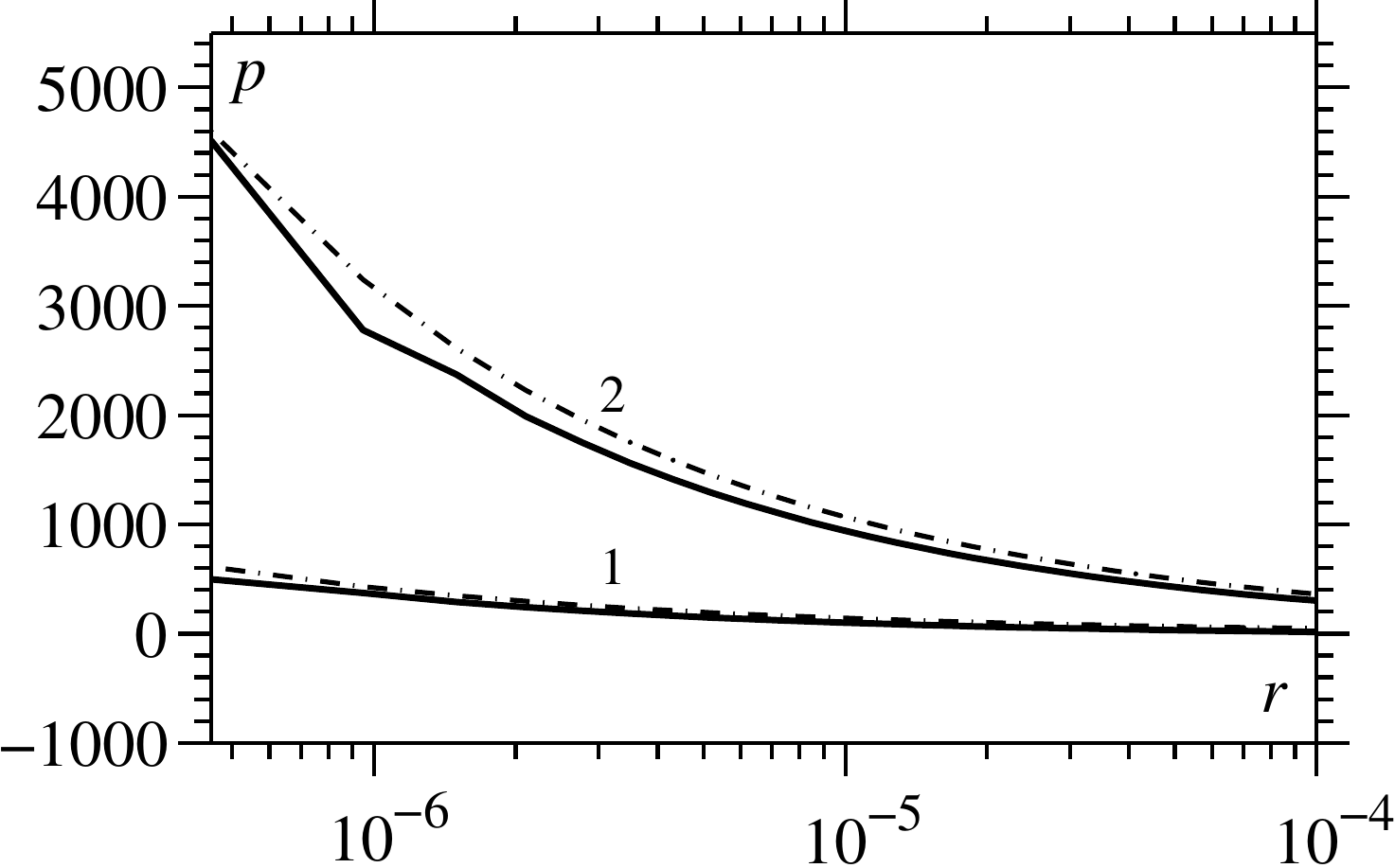} 
\includegraphics*[scale=0.55]{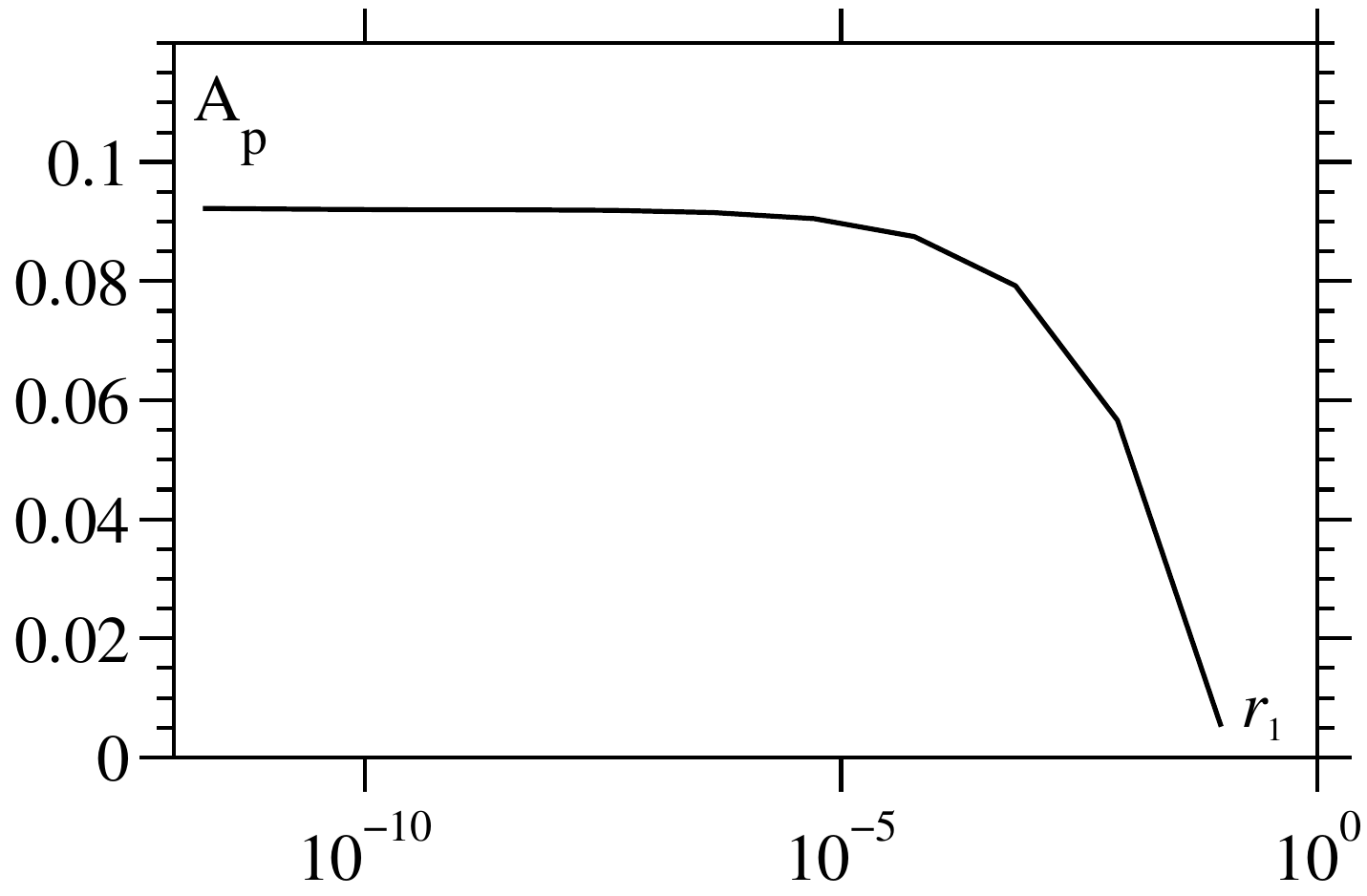} 
\caption{Left: computed pressure distributions in the vicinity of the corner along the liquid-solid
(1) and liquid-fluid (2) interfaces compared with the semi-analytic result (dashed lines). Right:
convergence of pressure coefficient $A_p$ to 0.092, plotted against $r_{1}$ the size of the
smallest element. }\label{F:175pressure_sing}
\end{figure}

In Fig.~\ref{F:175pressure_sing} with $\alpha=175^\circ$ and hence, from (\ref{lambda_eqn}),
$\lambda=1.529$, we see that the implementation of the new singular elements solves our previous
problems by (i) removing the oscillations in pressure as the corner point is approached, and (ii)
converging as the mesh is refined.

The value of $A=A_{p}$, which is the coefficient of the singular basis functions and is determined
by the finite element method, may be used after computation with the supplementary asymptotics of
Section~\ref{eigen} to produce a fully determined asymptotic solution for the velocity and
pressure. Then we may compare the analytic results of Section~\ref{eigen} with our numerical
results globally.  It should be pointed out that using the value of $A$ to extrapolate the analytic
behaviour of the eigensolution well outside the first elements provides a quick check to see if $A$
is in the correct range, this value does not in any way actually determine the velocity field
outside the first elements.  The comparison with pressure in Fig.~\ref{F:175pressure_sing} shows
good agreement between numerics and asymptotics; however, very close to the corner point along the
free surface the numerical solution is not as smooth as the asymptotic result. This is no surprise
given the huge gradients in pressure which are being approximated by linear basis functions both in
the radial and angular directions.

\begin{figure}
\centering
\includegraphics*[scale=0.55]{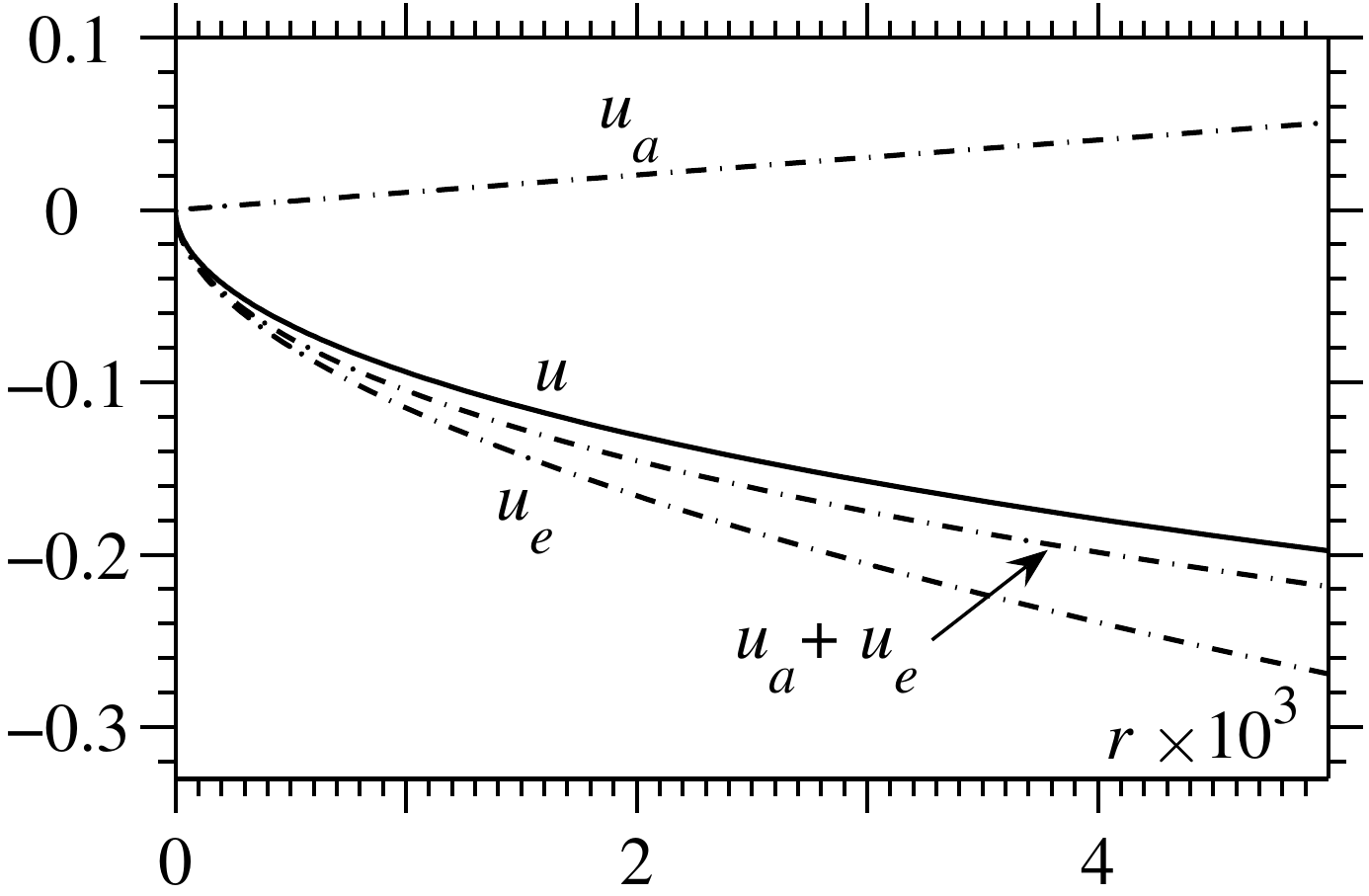}
\caption{Comparison of the computed velocity along the liquid-fluid interface $u$ with the
semi-analytic result which is shown decomposed into a contribution from the supplementary
asymptotics $u_a$ and the eigensolution contribution $u_e$ with A=0.092.}\label{F:175velocity_sing}
\end{figure}

In Fig.~\ref{F:175velocity_sing}, we compare the velocity along the interfaces of the computed
numerical solution to the asymptotic result, showing in particular how the full asymptotic solution
is a superposition of the eigensolution $u_{e}$ and the supplementary solution $u_{a}$. We see
that, although the supplementary asymptotics predicts that flow will be reversed near the contact
line, this is blown away by the strength of the eigensolution, which restores what one would
intuitively think is the correct direction for the flow.  The agreement we see in this figure is
sufficient when we consider that it is determined by the coefficient of the singular pressure which
only plays a role in elements adjacent to the corner point.

\subsubsection{Numerics incorporating two-term asymptotics}

For an angle of $\alpha=175^\circ$ we have completely resolved the situation.  However, we have
observed for smaller angles, roughly $\alpha_c<\alpha<150^\circ$, that, in terms of mesh
refinement, convergence is very slow, i.e. a mesh-independent regime is only realised for
exceptionally well resolved meshes which are well outside the scope of most numerical platforms
where the corner is just one part of a larger problem. In this range of angles, although
asymptotically in the limit $r\rightarrow0$ the logarithmic pressure behaviour ($p\sim\ln r$) is
overshadowed by the eigensolution, where $p\sim r^{\lambda-2}g(\theta)$, in reality, unavoidable
finiteness of the resolution of the mesh means that one could be sufficiently far away from the
corner point for the logarithm to be still dominant in the numerics. To see if this is indeed the
case, we consider the relative size of the two pressure terms at the edge opposite the corner point
in the first elements. Taking this element to have size $10^{-n}$, we see that the two singular
functions are comparable, at a critical value $\lambda_{c}$, at the edge of the first element when
\begin{equation}
\ln 10^{-n} = 10^{-n(\lambda_{c}-2)},
\end{equation}
that is when
\begin{equation}\label{lambda_critical}
\lambda_{c} =2-\frac{ \ln |-n\ln10| }{n\ln10}.
\end{equation}
Taking $n=6$, from (\ref{lambda_critical}) we have that $\lambda_{c} = 1.81$, which means, using
(\ref{lambda_eqn}), that the logarithmic behaviour dominates a numerical scheme, with the stated
spatial resolution, in the range of angles $\alpha_{c}<\alpha<143^\circ$. Thus, we have seen both
numerically and analytically, that even for an extremely well resolved mesh, we should expect that
there is a range of angles in which the logarithmic solution cannot be neglected; approximately for
$\alpha<150^\circ$. This is a very serious issue which must be resolved as almost all numerical
schemes will not be able to afford the required resolution to accurately capture the singular
behaviour.

The solution to this issue is to include the second term in the asymptotic expansion of pressure,
so that $p= A_{p}\Phi_p r^{\lambda-2} g(\theta)+B_{p}\Phi_p\ln r$ for $\alpha>\alpha_c$. The
coefficient to this logarithm $B_{p}$ is known exactly from the supplementary asymptotics
(\ref{pressure}) and we have already shown in Section~\ref{75} that this value is numerically
reproduced. Therefore, the simplest solution is to prescribe the value of $B_{p}$ and see if this
improves the speed of convergence of $A_{p}$.

\begin{figure}
\centering
\includegraphics*[scale=0.6]{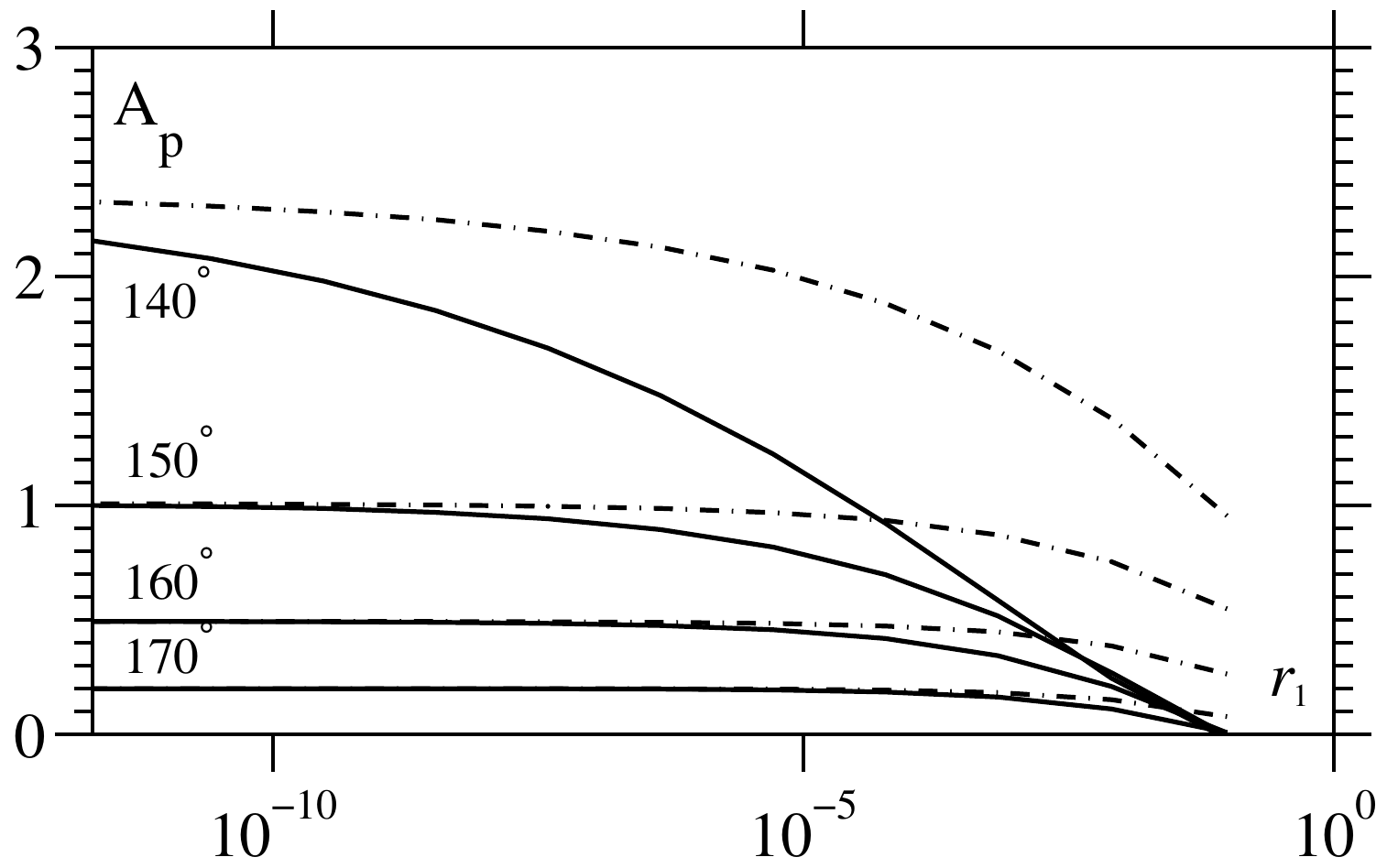}
\caption{A comparison of the convergence of the pressure coefficient
$A_{p}$ with mesh size
$r_{1}$. With (dashed line) and without (full line) the
supplementary asymptotics prescribed, for a
range of different angles.}\label{F:175_spedup}
\end{figure}

In Fig.~\ref{F:175_spedup} we show how nicely this method works with, as one would expect, the most
improvement occurring for smaller angles where the logarithmic pressure is strongest.  For example,
for $\alpha=150^\circ$, where the converged value is $A_{p} = 1.007$, if we neglect the logarithm,
then we require $\simeq 5000$ elements to get within $5\%$ of this value, but, if we include the
asymptotic logarithmic behaviour then we only need $\simeq 2500$ elements to attain the same
accuracy. Computationally this is a terrific saving and, given the simplicity of its
implementation, we conclude that the logarithmic asymptotic behaviour should be included for all
angles $\alpha>\alpha_c$.

\section{Conclusion}

We have shown that accurate numerical calculation of viscous flows near corners of the flow domain
in the framework of the finite-element method requires special treatment of the elements adjacent
to the corners. In the case of zero-stress/prescribed velocity boundary conditions on the sides of
the corner, the range of corner angles is split into two distinct regions. For the angles below the
critical angle $\alpha_c\approx128.7$, it is sufficient to use a logarithmic basis function for the
pressure, whereas for supercritical angles there appears a `hidden' eigensolution which
considerably complicates numerics. In order to obtain an acceptable (i.e.\ mesh-independent)
solution, one has to incorporate this eigensolution into the code by altering the basis functions
for the pressure in the elements adjacent to the corner point. Close to the critical angle it
becomes necessary to use a two-term asymptotics in the numerical algorithm, including the leading
terms of both the eigensolution and the solution of the inhomogeneous problem, as the finiteness of
the mesh size could result in the algebraically and logarithmically singular terms having
comparable values.

An issue that is now opened up for numerical and analytic investigation is how to generalize the
developed methods for a three-dimensional case, i.e.\ in a situation where both the contact angle
and the direction of velocity of the solid vary along a contact line. The first of these aspects
becomes particularly challenging when the angle varies from subcritical to supercritical.

\section*{Acknowledgements}
The authors kindly acknowledge the financial support of Kodak European Research and the EPSRC via a
Mathematics CASE award.

\bibliography{manuscript_arxiv}

\end{document}